\documentclass[aps,prl,showpacs,twocolumn,amsmath,amssymb,superscriptaddress]{revtex4-1}
\usepackage{graphicx}
\usepackage{bm}
\usepackage{mathptmx}
\usepackage[english]{babel}
\usepackage{indentfirst}
\usepackage{color}
\usepackage{listings}
\usepackage{multibib}
\newcommand{\kv}{\ensuremath{\mathbf{k}}}
\newcommand{\KV}{\ensuremath{\mathbf{q}}}
\newcommand{\QV}{\ensuremath{\mathbf{Q}}}

\begin{document}

\title{Fermi Condensation near van Hove Singularities within the Hubbard model on the Triangular Lattice}

\author{Dmitry Yudin}
\affiliation{Department of Physics and Astronomy, Uppsala University, Box 516, SE-751 20 Uppsala, Sweden}

\author{Daniel Hirschmeier}
\affiliation{Institut f\"ur Theoretische Physik, Universit\"at Hamburg, Jungiusstra\ss e 9, D-20355 Hamburg, Germany}

\author{Hartmut Hafermann}
\affiliation{Institut de Physique Th\'eorique (IPhT), CEA, CNRS, 91191 Gif-sur-Yvette, France}

\author{Olle Eriksson}
\affiliation{Department of Physics and Astronomy, Uppsala University, Box 516, SE-751 20 Uppsala, Sweden}

\author{Alexander I. Lichtenstein}
\affiliation{Institut f\"ur Theoretische Physik, Universit\"at Hamburg, Jungiusstra\ss e 9, D-20355 Hamburg, Germany}

\author{Mikhail I. Katsnelson}
\affiliation{Radboud University Nijmegen, Institute for Molecules and Materials, Heijendaalseweg 135, NL-6525 AJ Nijmegen, The Netherlands}
\affiliation{Department of Theoretical Physics and Applied Mathematics, Ural Federal University, Mira Street 19, 620002 Ekaterinburg, Russia}

\begin{abstract}
The proximity of the Fermi surface to van Hove singularities drastically enhances interaction effects and leads to essentially new physics. In this work we address the formation of flat bands (``Fermi condensation'') within the Hubbard model on the triangular lattice and provide a detailed analysis from an analytical and numerical perspective. To describe the effect we consider both weak-coupling and strong-coupling approaches, namely the renormalization group and dual fermion methods.
It is shown that the band flattening is driven by correlations and is well pronounced even at sufficiently high temperatures, of the order of 0.1--0.2 of the hopping parameter. The effect can therefore be probed in experiments with ultracold fermions in optical lattices.
\end{abstract}

\maketitle

{\it Introduction.---}The study of two-dimensional lattice models can potentially unveil the nature of exotic materials like unconventional superconductors and quantum spin liquids. After their almost simultaneous discovery, high-temperature superconductivity in cuprates \cite{Anderson,Scalapino,Dagotto} and the fractional quantum Hall effect \cite{Laughlin,Jain} posed some awkward questions to Landau Fermi liquid theory. For both systems, the Coulomb interaction is sufficiently strong to cause the breakdown of perturbative expansions. In such cases, the concept of quasiparticles providing a basis for understanding most of condensed-matter phenomena is questionable, and new physics can arise. In cuprates, the large onsite Coulomb repulsion eliminates the double occupancy and changes the statistics of charge carriers, while in the quantum Hall phase it leads to the formation of composite fermions. Both scenarios manifest deviations from Landau Fermi liquid behavior.

It is well known that many body effects are drastically enhanced in the vicinity of anomalies in the single-particle spectrum \cite{KT1985,Dzyaloshinskii1987,KT1990,HKV2011}. Soon after high-temperature superconductivity was detected in cuprates, it was pointed out that for the optimal doping the Fermi level lies in the vicinity of van Hove singularities (VHSs) with divergent density of states (DOS), and that in this case the Fermi liquid picture can be violated even for a weak interaction, due to singularities of the electron-electron vertex \cite{Dzyaloshinskii1987}. The concept of the so-called van Hove scenario has been pushed forward to explain a variety of phases associated with the presence of VHSs, e.g., superconductivity, itinerant ferromagnetism, and density waves. If the VHS is near the Fermi-level, both antiferromagnetism and $d$-wave superconductivity can be produced even at small on-site Coulomb repulsion, as can be shown from a renormalization group (RG) analysis \cite{Dzyaloshinskii1997,Schulz1997,Metzner2000,Honerkamp2001,Katsnelson2001,Igoshev2011} or the parquet approximation~\cite{Furukawa1998,Katsnelson2001}.
The nature of exotic ground states is determined by the delicate interplay of these fluctuations, which therefore remain controversial.

Ultracold Fermi gases in optical lattices \cite{Bloch,Lewenstein} open up completely new opportunities to study exotic states of interacting fermions. Today, the experimental realization of quantum many-body Hamiltonians, such as the Hubbard model, is a reality and a variety of system parameters such as the hopping, lattice type, and Hubbard repulsion can be tuned \cite{Bloch}. However, despite substantial progress in cooling, the achieved temperatures are still relatively high compared to the effective hopping parameter, so critical temperatures of the low-temperature phases cannot be reached. It is therefore important to identify effects that can be probed at these temperatures.

In this Letter, we show that a \emph{precursor} of a strongly correlated low-temperature instability, possibly chiral superconductivity \cite{Honerkamp2003}, exists at sufficiently high temperatures and that it can be probed in the paramagnetic phase of fermionic cold atoms on a triangular lattice. The effect can be understood in terms of {\it Fermi condensation}.

To clarify this statement, recall that in conventional Landau Fermi liquid theory \cite{Landau}, the free energy is a functional of the quasiparticle distribution function $n_{\kv}$. The particle distribution minimizes this functional, i.e., $\delta F[n_{\kv}]/\delta n_{\kv}=0$, which leads to
\begin{gather}
\label{var}
\varepsilon_{\kv}(T)=\mu(T)+T\log\left[\left(1-n_{\kv}\right)/n_{\kv}\right],
\end{gather}
where $\varepsilon_{\kv}$ is the dispersion, $\mu$ the chemical potential, and $T$ denotes temperature.
This expression reproduces the celebrated Fermi distribution $n_{\kv}=1/(1+e^{(\varepsilon_{\kv}-\mu)/T})$. On the other hand, $\varepsilon_{\kv}(T)$ is a functional of $n_{\kv}$. As long as the group velocity is positive, all variations $\delta E$ of this functional are positive and the Fermi distribution corresponds to the minimum. If the group velocity of the quasiparticles becomes negative, there exist variations for which $\delta E<0$. This leads to a restructuring of the distribution function in a certain interval of momenta $k_{i}<k<k_{f}$, where the resulting $n_{\kv}$ differs from the Fermi distribution, but still minimizes the functional.
In the limit $T\to 0$, $\varepsilon_{\kv}=\mu$ and hence the dispersion becomes entirely flat in this interval. In analogy to the Bose-Einstein condensate, this highly degenerate state has been termed Fermi condensation.

The idea was suggested~\cite{Khodel,Zverev} in a purely phenomenological background and remains controversial~\cite{Nozieres}.  If it exists, a Fermi condensate is a new state of matter which is topologically different from both the Fermi liquid and the Luttinger liquid~\cite{Volovik}. In the context of the van Hove scenario in high-temperature superconductivity, the Fermi condensation was considered in Ref.~\cite{IKK2002} as a way to demonstrate that the van Hove scenario is not just a scenario at van Hove filling and hence for a single point (an objection from Ref. \cite{Anderson}); because of the formation of flat bands, there is a pinning of the Fermi energy to the VHS point for a whole range of electron concentrations.
Otherwise, below a critical temperature, the highly degenerate state may give way to another fermionic instability associated with a non-Fermi liquid ground state. It is therefore important to observe this precursor effect experimentally.

We address this effect for the Hubbard model at triangular lattice from both weak-coupling and strong-coupling limits, by means of RG and dual-fermion \cite{dual1} approaches, respectively. Our analysis shows that the phenomenon is robust and can be observed in experiments with ultracold Fermi gases at sufficiently high temperatures.

\begin{figure}[h]
\begin{center}
\includegraphics[scale=1.25]{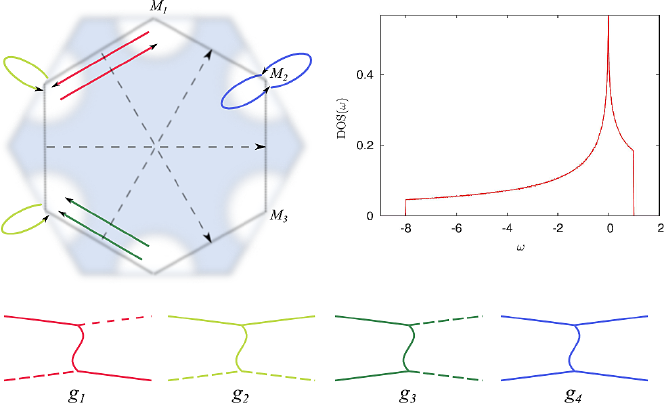}
\caption{\label{brzone}(Color online) Hexagon-shaped Brillouin zone and DOS of the system doped to the VHS. From momenta and spin conservation the following two-particle processes are allowed: exchange scattering ($g_1$), forward scattering ($g_2$), umklapp scattering ($g_3$), and intrapatch scattering ($g_4$).}
\end{center}
\end{figure}

\emph{Model.---}We focus on a Hubbard model on the triangular lattice,
\begin{equation}
H=\sum\limits_{\mathbf{k}\sigma}\varepsilon_\mathbf{k}d_{\mathbf{k}\sigma}^\dagger d_{\mathbf{k}\sigma}+U\sum\limits_{i}n_{i\uparrow}n_{i\downarrow}
\end{equation}
with local Coulomb repulsion $U>0$ and dispersion relation $\varepsilon_\mathbf{k}=-2t[\cos(k_xa)+2\cos(k_xa/2)\cos(k_ya\sqrt{3}/2)]-\mu$, 
where $t>0$ is the hopping amplitude, $\mu$ the chemical potential, and $a$ is the lattice spacing. We take $a=1$ in the following.
The reciprocal lattice is spanned by the vectors $\mathbf{G}_1=2\pi\left(\mathbf{e}_x\sqrt{3}-\mathbf{e}_y\right)/\sqrt{3}$ and $\mathbf{G}_2=4\pi\mathbf{e}_y/\sqrt{3}$, while the first Brillouin zone is hexagon shaped. At $3/4$ filling, logarithmic VHSs (kinks in the DOS) appear in three inequivalent saddle points $M_1=\left(0,2\pi/\sqrt{3}\right)$, $M_{2,3}=\left(\pi,\pm\pi/\sqrt{3}\right)$,  and the hexagon-shaped Fermi surface becomes highly nested (Fig. \ref{brzone}).
It is well known that in the weak coupling limit $U/t\ll 1$, the dominant instability for a non-nested Fermi surface away from VHSs is related to superconductivity. Contrary to this, at VHSs ($\nabla_\mathbf{k}\varepsilon_\mathbf{k}=0$) the Fermi surface has flat sides and is nested as a result. The vector $\QV_{\alpha\beta}$ connecting different points $M_\alpha$ and $M_\beta$ is such that $2\QV_{\alpha\beta}=0$ modulo a reciprocal lattice vector.
In what follows we will focus on the model doped exactly to the VHS ($\mu=2t$) and perfect nesting.

{\it Weak-coupling analysis.---}We start our analysis of the RG flow by developing a three-patch renormalization group analogous to Refs. \cite{IKK2002,Katanin2003}.
The number of patches for the triangular lattice agrees with the number of inequivalent saddle points, in which the DOS diverges logarithmically: $N=N_0\log\left[\Lambda/\max\left(2t,T\right)\right]$ (here $\Lambda$ is a high-energy cutoff). The problem in question can be reduced to a quasi-one-dimensional one if we introduce those two-particle scattering processes between different patches, which are allowed by momentum conservation. One-dimensional systems are known to be unstable to the formation of pair instabilities in both Cooper (particle-particle) and Peierls (particle-hole) channels, and result in logarithmic singularities for pair susceptibilities. Extending the quasi-one-dimensional analysis we define four different interactions associated with two-particle scattering between different patches: exchange (or backward) scattering ($g_1$), forward scattering ($g_2$), umklapp scattering ($g_3$), which conserves momentum modulo a reciprocal lattice vector, and intrapatch scattering ($g_4$). All four interactions are marginal at tree level, but acquire logarithmic corrections from the integration near the VHS, thus justifying the use of logarithmic RG. These logarithmic corrections come from energy scales $E<\Lambda\approx t$, the energy scale at which higher-order corrections to the dispersion become important.

\begin{figure}[h]
\begin{center}
\includegraphics[scale=0.33]{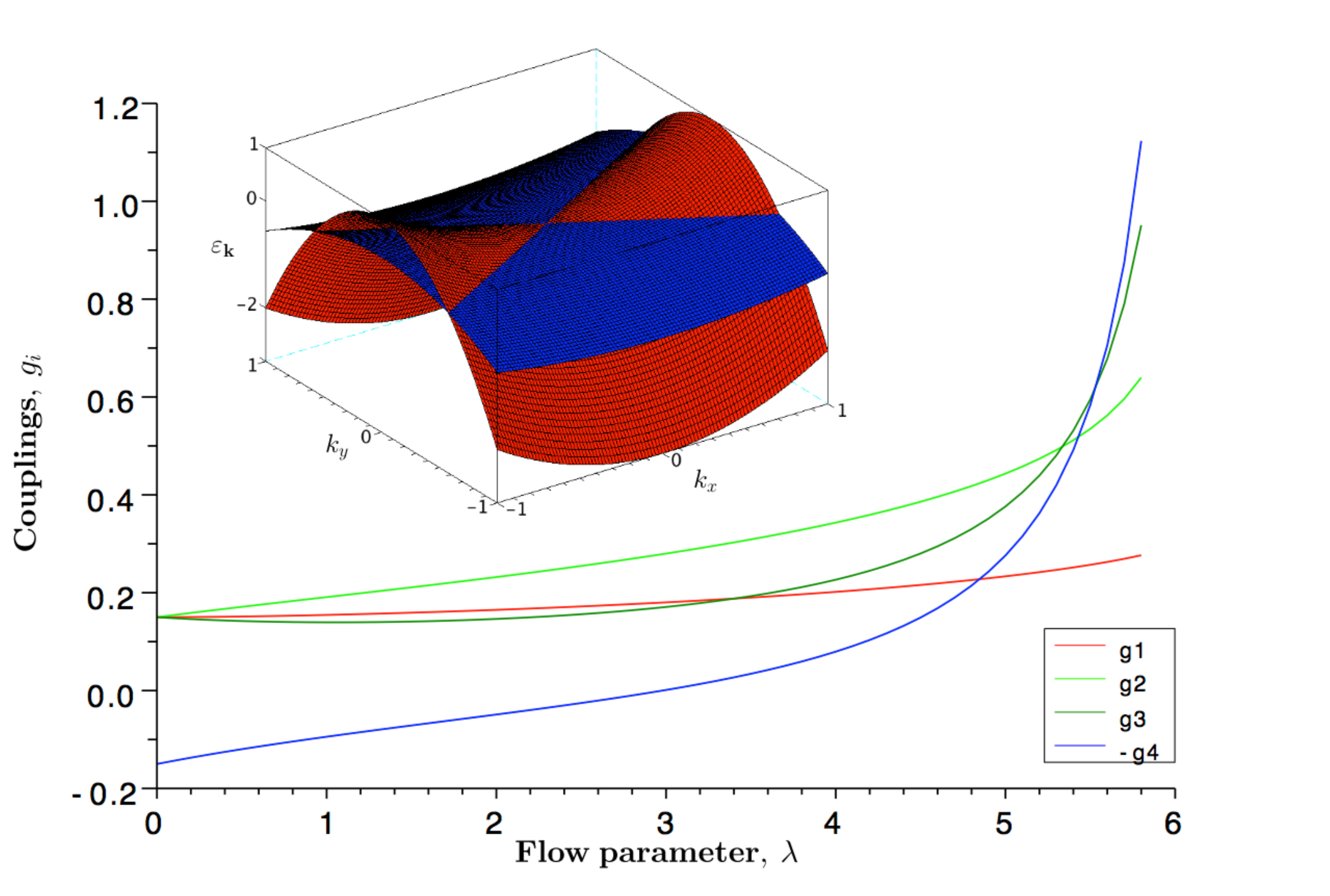}
\caption{\label{renormalization} (Color online) Main panel: Renormalization group flow of the couplings $g_{i}$.
Inset: Dispersion relation in the vicinity of the saddle point corresponding to the bare (red) and renormalized (blue) action. The flattening of the band is clearly visible. The plotting region is determined by the cutoff parameter $\Lambda/t\sim1$.}
\end{center}
\end{figure}

The susceptibilities in the particle-particle $\chi_{pp}(\KV=\mathbf{0})=N_0\log\left[\Lambda/\max\left(2t,T\right)\right]\log(\Lambda/T)/2$ and particle-hole $\chi_{ph}(\KV=\QV_{\alpha\beta})=N_0\log^2\left[\Lambda/\max\left(2t,T\right)\right]/2$ channels, evaluated at momentum transfers zero and $\QV_{\alpha\beta}$ between points $M_\alpha$ and $M_\beta$, are log-square divergent. One logarithm stems from the DOS, whereas the second is inherent to the divergence in the Cooper channel for $\chi_{pp}$ and appears in $\chi_{ph}(\QV)$ due to perfect nesting of the Fermi surface. For the analysis of the low-energy properties we neglect the logarithmically divergent contributions $\chi_{ph}(\mathbf{0})$ and $\chi_{pp}(\QV)$, which are parametrically smaller. Restricting the integration region to the patches and placing external momenta at the critical points, we derive one-loop RG equations using momentum-shell integration \cite{sup} with respect to the flow parameter $\lambda=\chi_{pp}(\KV=\mathbf{0},E)$. It is noteworthy that to leading order the solution to a set of RG equations is defined by the relative weight between the Peierls and Cooper channels only. Because of nesting the flow of the coupling constants is strongly modified and the effect of interactions is dramatically enhanced. An inspection of RG flow in Fig. \ref{renormalization} reveals that the couplings diverge when approaching instability region $\lambda_c$ with $|g_4|>g_3>g_2>g_1$; i.e., intrapatch scattering prevails. The combination of a divergent DOS and perfect nesting leads to a RG flow to strong coupling, in agreement with an earlier fRG study \cite{Honerkamp2003}. Thus, the local repulsive coupling can favor the formation of instabilities towards magnetic or superconducting states at relatively high temperatures  $\lambda_c=\chi_{pp}(E=T_c)$, e.g., for the initial values of running couplings $g_0$,

\begin{align}\label{temp}
T_c\sim t\exp\left(-1/\sqrt{g_0N_0}\right)
\end{align}

\noindent  even if the interaction strength is weak compared to the fermionic bandwidth $W$.

In order to obtain the renormalized band function we proceed by estimating the second-order correction to the self-energy $\Sigma_\omega(\mathbf{k})$ for $\mathbf{k}$ near $M_1$. Similar to \cite{IK2001,IKK2002} we make a distinction among three contributions stemming from intermediate integration with quasimomentum corresponding to the same point and the two other VHSs: $\Sigma_\omega(\mathbf{k})=\sum_{i=1,2,3}\Sigma^i_\omega(\mathbf{k})$ \cite{sup}. The band function is determined by the pole of the cutoff-independent Green's function that can be obtained by solving the corresponding Dyson equation, whereas the effects of spectrum renormalization, which describe the flattening, can be absorbed into mass renormalization factors. The remaining divergencies are to be associated with the quasiparticle residue. The resulting quasiparticle spectrum in the vicinity of the $M$ point (with initial $g_1=g_2=g_3=g_4=0.15$) is shown in the inset of Fig.~\ref{renormalization}: The spectrum is almost flat in a rather wide range of $\mathbf{k}$ resulting from mass renormalization. The quasiparticle weight is also renormalized under the RG flow (not shown). We find that the pinning of the Fermi level to the VHS remains robust under the RG flow. Thus, we conclude that the effects of renormalization drastically affect the Fermi surface topology, leading to the formation of an extended VHS.

{\it Strong-coupling analysis.---}In order to demonstrate the robustness and experimental accessibility of the phenomenon, it is necessary to show that the effect persists at finite temperatures and strong interaction. This is a challenging task: While dynamical mean-field theory (DMFT) captures nonperturbative phenomena such as the Mott transition, it neglects spatial correlations. Because of the important role of susceptibilities, the problem cannot be treated in DMFT. Cluster extensions of DMFT \cite{cDMFT} lack sufficient momentum resolution. Both criteria are met only in novel approaches combining DMFT with analytical methods \cite{DGA,dual1}. Here we employ the {\it dual fermion} technique \cite{dual1} (see \cite{Hart2010} for a comprehensive overview).

In this approach, the electronic self-energy is decomposed into a local part obtained from DMFT and a \emph{nonlocal} momentum dependent correction $\Sigma_{\omega}(\kv) = \Sigma^{\text{DMFT}}_{\omega} + \Sigma^{\text{NL}}_{\omega}(\kv)$, which is evaluated in dual perturbation theory.
The antiferromagnetic pseudogap, Fermi-arc formation, and non-Fermi-liquid effects due to the VHS are already captured by the lowest-order diagrams \cite{dual3}. Here we employ the ladder approximation, which describes the feedback of collective excitations on the electronic self-energy. Introducing the dual particle-hole bubble $\tilde{\chi}^\nu_{\omega}(\KV)=-T\sum_{\kv}\tilde{G}_{\omega}(\kv)\tilde{G}_{\omega+\nu}(\kv+\KV)$, the dual self-energy reads
\begin{align}
\label{sigma}
\tilde{\Sigma}_{\omega}(\kv) = T\!\!\!\sum_{\KV\omega'\nu} \gamma^\nu_{\omega\omega'} \tilde{G}_{\omega+\nu}(\kv+\KV)\tilde{\chi}^\nu_{\omega'}(\KV)[\Gamma^\nu_{\omega'\omega}(\KV)-\frac{1}{2}\gamma^\nu_{\omega'\omega}].
\end{align}
Here $\gamma^\nu_{\omega\omega'}$ is the fully connected dynamical vertex of the impurity model \cite{Hart2010}, $\omega$, $\nu$ denote fermionic and bosonic Matsubara frequencies, respectively, and $T$ denotes temperature. The vertices are tensors in spin space and spin summations have been omitted for clarity. The second term in brackets prevents double counting of diagrams. From the Bethe-Salpeter equation $[\Gamma_\nu^{-1}(\KV)]_{\omega\omega'} = [\gamma_\nu^{-1}]_{\omega\omega'} - \tilde{\chi}^\nu_{\omega}(\KV)\delta_{\omega\omega'}$ we obtain the vertex function $\Gamma$. The bare dual Green's function is $\tilde{G}^{0}_\omega(\kv)=G^{\text{DMFT}}_\omega(\kv)-g_\omega$, where $g_\omega$ is the exact local DMFT Green's function. This approximation is applicable for strong coupling \cite{LDFA}. The relation between the $\tilde{\Sigma}$ and the lattice Green's function can be written in the form \cite{dual3,Hart2010}
\begin{align}
\label{glat}
G_{\omega}(\kv)=[(g_{\omega}+g_{\omega} \tilde{\Sigma}_{\omega}(\kv)  g_{\omega} )^{-1}+\Delta _{\omega } -\varepsilon_\mathbf{k} ]^{-1},
\end{align}
with the DMFT hybridization function $\Delta_{\omega}$.
\begin{figure}[b]
\begin{center}
\includegraphics[scale=.7]{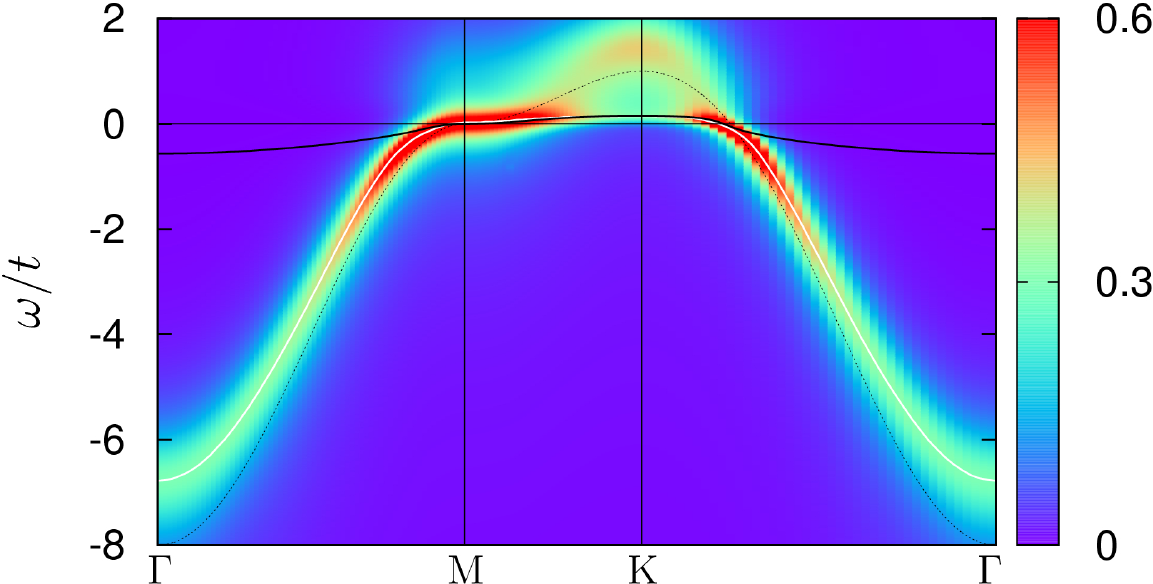}
\caption{\label{spectral} (Color online) Spectral function in dual fermion approach at $U/t=8$ and $T/t=0.05$. Local maxima corresponding to the lower band are indicated by a white line. In the vicinity of the Fermi level, this lower band perfectly matches the prediction $\epsilon_{\kv}-\mu=T\ln [(1-n_{\kv})/n_{\kv}]$ following from the Fermi condensate hypothesis (thick black line). The bare dispersion is shown for comparison (blue, dashed).}
\end{center}
\end{figure}

The resulting spectral function $-(1/\pi)\mathrm{Im}G_{\omega}(\kv) $ for $U/t=8$ is shown in Fig.~\ref{spectral}. We observe a broadening and flattening of the spectrum at the $M$ point. While flattening of the spectrum is partly present in DMFT due to band renormalization, including spatial correlations leads to the formation of an extended VHS.
Apart from the incoherent high-energy excitations we observe a well-defined and only slightly dispersive band at low energies, which spans a large region of the Brillouin zone between the $M$ and $K$ points. We have marked the local maxima with a white line. 
We find that this band agrees perfectly well with the prediction $\epsilon_{\kv}-\mu=T\ln[(1-n_{\kv})/n_{\kv}]$ from Eq. \eqref{var} (black line) everywhere in the vicinity of the Fermi level. While the results are described by the Landau functional, the self-energy clearly exhibits a power law and hence non-Fermi liquid behavior. 
For $T\to  0$ this leads to a flat band and Fermi condensation, or the system becomes unstable due to the degeneracy. We therefore interpret the effect as a precursor to a correlated magnetic or superconducting ground state.
The formation of this band is correlation driven as it disappears when the interaction is lowered.

In order to further elucidate the origin of this effect, we note that because of the large DOS at the $M$ point due to the proximity of the VHS, the dominating contribution to the convolution in the self-energy \eqref{sigma} in the vicinity of the $M$ point is expected to stem from the vicinity of the $\Gamma$ point. An analysis of the leading eigenvalues of the Bethe-Salpeter equation reveals that the spin channel dominates in the vicinity of $\Gamma$ in agreement with our RG analysis, where intrapatch scattering is found to give the dominant contribution. Hence the effect results from the combination of a large DOS and coupling to strong ferromagnetic spin fluctuations. Indeed, our calculations unambiguously determine this effect to originate from collective excitations in the spin channel \cite{sup}. The observed tendency to ferromagnetic ordering due to frustration is in line with previous results \cite{dualtriang}.

The large self-energy in the vicinity of the $M$ point leads to both a broadening of the spectrum and a strong reduction of spectral weight at the $M$ point, also in agreement with the RG.
The flattening is considerably stronger in non-self-consistent calculations, where attenuation of the fluctuations due to damping of quasiparticles at the $M$ point is not taken into account \cite{sup}. The absence of the low-energy band in second-order approximation to the dual self-energy underlines the importance of the feedback of collective excitations onto the electronic degrees of freedom.

\begin{figure}[t]
\begin{center}
\includegraphics[scale=1.]{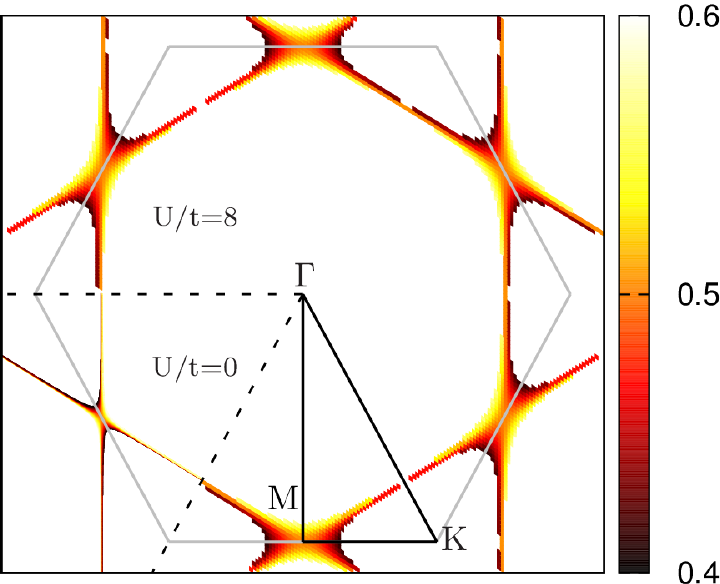}
\caption{\label{fsnk} (Color online) Broadened Fermi surface within $\pm 0.1$ electrons for $U/t=8$ and $T/t=0.1$. The lower left sextant shows the noninteracting result.}
\label{fs}
\end{center}
\end{figure}

In the top panel of Fig. \ref{fsnk} we plot the so-called broadened Fermi surface within $\pm 0.1$ electrons from the value 0.5 corresponding to the interacting Fermi surface for given temperature. This quantity is directly related to the occupation function for different momenta, which is experimentally measurable~\cite{Bloch}.
The comparison with the noninteracting case shows that the effect of flattening is substantial. Increasing the interaction strength $U$ strongly enhances the flattening while lowering the temperature mitigates it. The correlation-driven effect can, nevertheless, clearly be separated from this purely thermal effect even at the highest temperatures (see Supplemental Material \cite{sup}).
We find that the effect persists up to shifts in chemical potential of at least $0.5t$, showing that it is robust to the presence of a trapping potential.

{\it Conclusions.---}In summary, we have investigated the formation of extended van Hove singularities in the triangular lattice. The renormalization group and strong-coupling numerical analysis establish the phenomenon as driven by many-body interactions: The interplay of many-particle scattering and nesting leads to band flattening near van Hove singularities.
The related high intensity in the spectral function may find interesting applications in tunneling experiments or spintronics.
The phenomenon can be interpreted as a precursor to a strongly correlated many-body ground state. Its study in the controlled environment of cold atom experiments  may fundamentally improve our understanding of correlated systems.
We have shown the effect to be robust when tuning interaction, temperature, and chemical potential. In particular, its signature in the occupation function is found to persist to relatively high temperatures, making the phenomenon detectable in experiments with ultracold atoms in optical lattices. The flat band could be observed directly via band spectroscopy~\cite{Heinze} or indirectly via the momentum distribution function accessible in time-of-flight measurements~\cite{Bloch}.

We thank K. Sengstock, Ch. Becker, and A. Rubtsov for fruitful
discussions. We further acknowledge support from the Deutsche
Forschungsgemeinschaft (SFB925), the CUI Cluster of Excellence, and computer support from the NIC, Forschungszentrum J\"ulich, under project HHH14.
M.I.K. acknowledges financial support from ERC (project 338957 FEMTO/NANO) and from NWO via the Spinoza Prize, O.E. acknowledges support from VR, the KAW Foundation and the ERC (project 247062---ASD), and H.H. acknowledges support from the FP7/ERC, under Grant No. 278472-MottMetals.
The dual fermion calculations used an adapted version of an open source hybridization expansion quantum Monte Carlo algorithm~\cite{cthyb} based on the ALPS libraries~\cite{ALPS}.

\onecolumngrid
\setcounter{page}{-1}
\setcounter{table}{0}
\setcounter{section}{0}
\setcounter{figure}{0}
\setcounter{equation}{0}
\renewcommand{\thepage}{\Roman{page}}
\renewcommand{\thesection}{S\arabic{section}}
\renewcommand{\thetable}{S\arabic{table}}
\renewcommand{\thefigure}{S\arabic{figure}}
\renewcommand{\theequation}{S\arabic{equation}}
\cleardoublepage
\vfill\eject
\thispagestyle{empty}
\phantom{hi}\vfill\eject
\section{Supplemental Material}
\section{Model}

We start our analysis with the Hubbard model on the triangular lattice. The band function of non-interacting electrons corresponding to nearest-neighbor hopping is

\begin{equation}\label{dispersion}
\varepsilon_\mathbf{k}=-2t\left(\cos k_x+2\cos\dfrac{k_x}{2}\cos\dfrac{k_y\sqrt{3}}{2}\right)-\mu,
\end{equation}

\noindent with the hopping parameter $t>0$ and the chemical potential $\mu$ respectively. We set the lattice spacing $a=1$. The dispersion relation possesses three inequivalent saddle points at

\begin{equation*}
M_1=\left(0,\dfrac{2\pi}{\sqrt{3}}\right), \quad M_{2,3}=\left(\pi,\pm\dfrac{\pi}{\sqrt{3}}\right)
\end{equation*}

\noindent which are known to coincide with the Van Hove singularities (VHSs) of the model in question. On being doped to $\mu=2t$ the density of states (DOS)

\begin{equation}\label{dos}
N(\varepsilon)=\dfrac{1}{\pi^2t\sqrt{3}}K\left(\dfrac{1}{2}+\dfrac{|\varepsilon|+2t-\varepsilon^2/4t}{4\sqrt{t\left(|\varepsilon|+t\right)}}\right)\sim_{|\varepsilon|/t\ll 1} N_0\log\left(\dfrac{2t}{|\varepsilon|}\right)
\end{equation}

\noindent diverges logarithmically, where $N_0=\sqrt{3}/(2\pi^2t)$ and $K(z)$ is the complete elliptic integral of the first kind \cite{GradRyzh_sup}.

We proceed by patching the hexagon-shaped Brillouin zone near $M_1$, $M_2$, and $M_3$. By restricting quasiparticle dynamics to these regions we can introduce the corresponding operators $a_1$, $a_2$, and $a_3$. The most general expression of the action that meets momentum (up to reciprocal lattice vector) and spin conservation reads as follows

\begin{gather}\nonumber
S[a_1,a_2,a_3]=\int\limits_0^\beta d\tau\sum\limits_{n\mathbf{k}\sigma}\bar{a}_{n\mathbf{k}\sigma}\left(\partial_\tau+\varepsilon_n\right)a_{n\mathbf{k}\sigma}+\dfrac{g_1}{2}\int\limits_0^\beta d\tau\sum\limits_{m\neq n}\sum\limits_{\{\mathbf{k}\}}\bar{a}_{m\mathbf{k}_1\sigma} \bar{a}_{n\mathbf{k}_2\sigma^\prime} a_{m\mathbf{k}_3\sigma^\prime}a_{n\mathbf{k}_4\sigma}+\dfrac{g_2}{2}\int\limits_0^\beta d\tau\sum\limits_{m\neq n}\sum\limits_{\{\mathbf{k}\}}\bar{a}_{m\mathbf{k}_1\sigma} \bar{a}_{n\mathbf{k}_2\sigma^\prime} a_{n\mathbf{k}_3\sigma^\prime}a_{m\mathbf{k}_4\sigma} \\ \label{action}
+\dfrac{g_3}{2}\int\limits_0^\beta d\tau\sum\limits_{m\neq n}\sum\limits_{\{\mathbf{k}\}}\bar{a}_{m\mathbf{k}_1\sigma} \bar{a}_{m\mathbf{k}_2\sigma^\prime} a_{n\mathbf{k}_3\sigma^\prime}a_{n\mathbf{k}_4\sigma}+\dfrac{g_4}{2}\int\limits_0^\beta d\tau\sum\limits_n\sum\limits_{\{\mathbf{k}\}}\bar{a}_{n\mathbf{k}_1\sigma} \bar{a}_{n\mathbf{k}_2\sigma^\prime} a_{n\mathbf{k}_3\sigma^\prime}a_{n\mathbf{k}_4\sigma}.
\end{gather}

\noindent It includes exchange scattering ($g_1$), forward scattering ($g_2$), umklapp scattering ($g_3$), and intrapatch scattering ($g_4$) (see the explanation in the main text). The indices $m$ and $n$ run over 1, 2, and 3, while $\sum\limits_{\{\mathbf{k}\}}$ stands for $\sum\limits_{\mathbf{k}_1+\mathbf{k}_2=\mathbf{k}_3+\mathbf{k}_4}\sum\limits_{\sigma\neq\sigma^\prime}$.

\section{Renormalization group analysis}

Applying the formalism of Wilsonian renormalization group to \eqref{action} leads to a set of renormalization group (RG) equations \cite{LeHur_sup}:

\begin{align}
\dfrac{dg_1}{d\lambda}&=2d_1g_1\left(g_2-g_1\right)+d_2g_1\left(g_1+2g_4\right)-2d_3g_1g_2, \\
\dfrac{dg_2}{d\lambda}&=d_1\left(g_2^2+g_3^2\right)+2d_2\left(g_1-g_2\right)\left(g_2+g_4\right)-d_3\left(g_1^2+g_2^2\right), \\
\dfrac{dg_3}{d\lambda}&=-g_3\left(g_3+2g_4\right)+2d_1g_3\left(2g_2-g_1\right), \\
\dfrac{dg_4}{d\lambda}&=-2g_3^2-g_4^2+d_2\left(2g_1^2+4g_1g_2-4g_2^2+g_4^2\right).
\end{align}

\noindent where the RG scale is chosen to be equal to $\lambda=\chi_{pp}(\KV=\mathbf{0})$, i.e. the susceptibility in Cooper channel at zero momentum transfer,

\begin{equation}\label{ppbubble}
\chi_{pp}(\mathbf{0})=T\sum\limits_\omega\int\dfrac{d^2k}{(2\pi)^2}G(i\omega,\mathbf{k})G(-i\omega,-\mathbf{k}).
\end{equation}

\noindent Inserting the Green function $G(i\omega,\mathbf{k})=\left(i\omega-\varepsilon_\mathbf{k}\right)^{-1}$ into (\ref{ppbubble}), with the band function $\varepsilon_\mathbf{k}$ expanded near the VHS and DOS (\ref{dos}) we arrive at

\begin{equation}
\chi_{pp}(\mathbf{0})=\dfrac{1}{2}\int\limits^\Lambda_{-\Lambda}\dfrac{d\varepsilon}{\varepsilon}N(\varepsilon)\tanh\left(\dfrac{\varepsilon}{2T}\right)=\dfrac{N_0}{2}\log\left(\dfrac{\Lambda}{\max\left(2t,T\right)}\right)\log\left(\dfrac{\Lambda}{T}\right)
\end{equation}

\noindent within logarithmic accuracy. The particle-particle susceptibility is doubly logarithmically divergent: the first logarithm comes from the DOS, whereas the second one takes its origin in the Cooper instability. Being doped to VHS the perfectly nested Fermi-surface drastically enhances the interaction effects in the particle-hole, or Peierls, channel. In fact, $\chi_{ph}(\QV)$ at momentum transfer $\mathbf{Q}$ (a vector connecting two different critical points, $\varepsilon_{\mathbf{k}+\mathbf{Q}}=-\varepsilon_\mathbf{k}$) can be estimated in a similar manner:

\begin{figure}[t]
\begin{center}
\includegraphics[scale=1.]{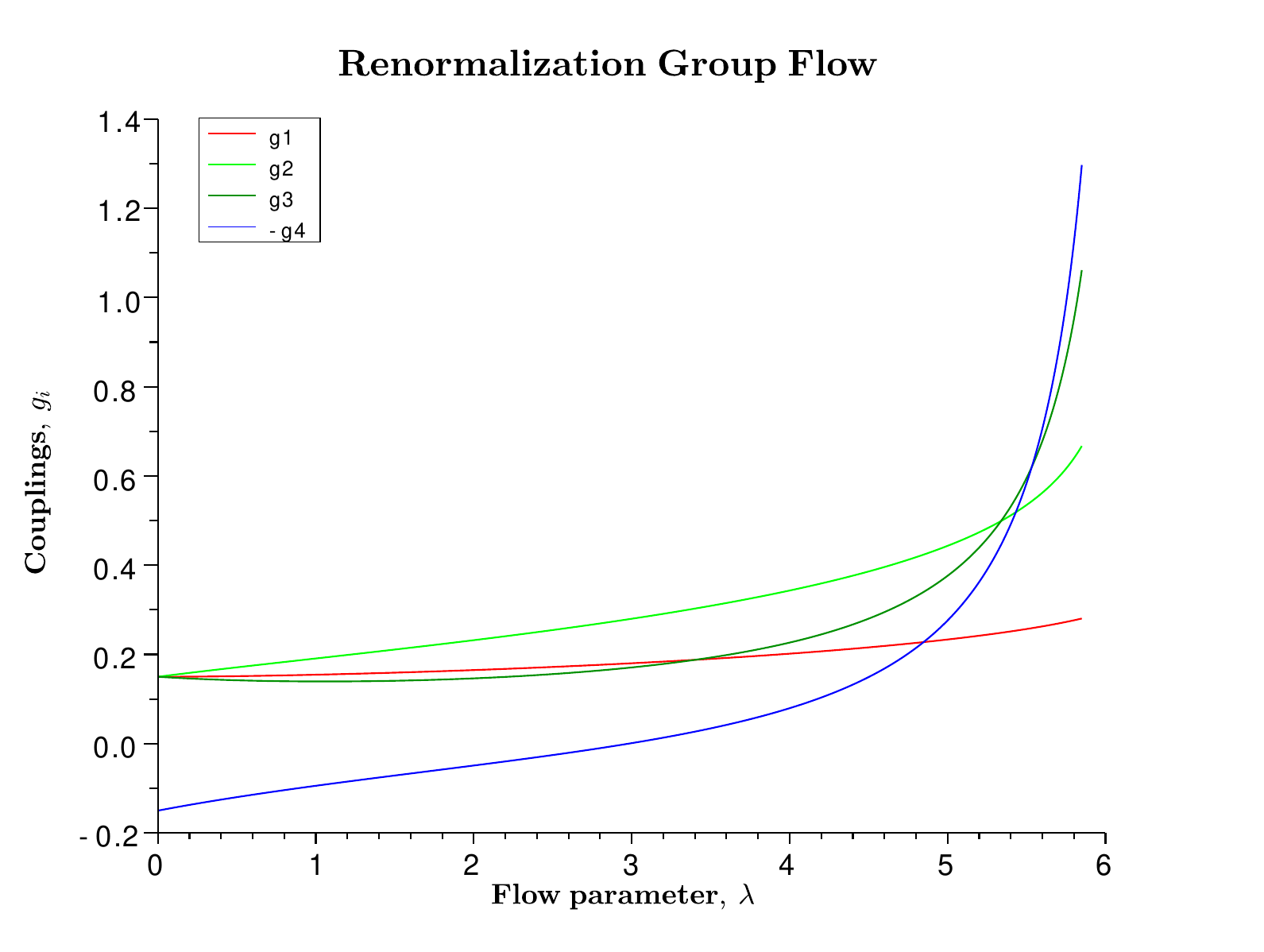}
\caption{The renormalization group flow.}\label{flow}
\end{center}
\end{figure}

\begin{equation}
\chi_{ph}(\QV)=-T\sum\limits_\omega\int\dfrac{d^2k}{(2\pi)^2}G(i\omega,\mathbf{k})G(i\omega,\mathbf{k}+\mathbf{Q})=\dfrac{N_0}{2}\log^2\left(\dfrac{\Lambda}{\max\left(2t,T\right)}\right).
\end{equation}

\noindent The latter does not hold away from perfect nesting. The coefficient $d_{1}$

\begin{equation}
d_1=\dfrac{d\chi_{ph}(\QV)}{d\lambda}\approx\dfrac{\chi_{ph}(\QV)}{\chi_{pp}(\mathbf{0})},
\end{equation}

\noindent showing the ratio between the Peierls and Cooper channels respectively smoothly interpolates between its asymptotic forms $d_1(\lambda=0)=1$ and $d_1(\lambda\gg1)\sim 1/\sqrt{\lambda}$, whereas the remaining coefficients, $d_2=d\chi_{ph}(\mathbf{0})/d\lambda$ and $d_3=-d\chi_{pp}(\QV)/d\lambda$ are parametrically smaller (sub-leading contribution). For the initial values $g_1=g_2=g_3=g_4=0.15$ pointed out in the text and the model function $d_1(\lambda)=1/\sqrt{1+\lambda}$ the RG flow is shown on Figure \ref{flow}.

\section{Spectrum renormalization}

In this section we provide a basic sketch of the self-energy computation to second order in interaction strength. As explained in the main text, we separate the contribution coming from the intermediate integration with quasimomentum from the same (in what follows $M_1$) and different patches. Near $M_1$ the bare band function reads $\varepsilon_\mathbf{k}=t\left(k_x^2-3\tilde{k}_y^2\right)/2$, with $\tilde{k}_y=k_y-2\pi/\sqrt{3}$, while the second order contribution to the self-energy can be evaluated as follows

\begin{equation}\label{selfenergy}
\Sigma(i\omega,\mathbf{k})=U^2\int\dfrac{d^2k_1}{(2\pi)^2}\int\dfrac{d^2k_2}{(2\pi)^2}\dfrac{f(\varepsilon_{\mathbf{k}_1})f(-\varepsilon_{\mathbf{k}_2})f(\varepsilon_{\mathbf{k}-\mathbf{k}_1+\mathbf{k}_2})+f(-\varepsilon_{\mathbf{k}_1})f(\varepsilon_{\mathbf{k}_2})f(-\varepsilon_{\mathbf{k}-\mathbf{k}_1+\mathbf{k}_2})}{i\omega-\varepsilon_{\mathbf{k}_1}+\varepsilon_{\mathbf{k}_2}-\varepsilon_{\mathbf{k}-\mathbf{k}_1+\mathbf{k}_2}}.
\end{equation}

\noindent We are interested in a small region close to the VHS, so that (\ref{selfenergy}) is to be represented by its Taylor expansion

\begin{equation}
\Sigma_\Lambda(i\omega,\mathbf{k})=\alpha(\Lambda)\left(i\omega\right)+\beta(\Lambda)k_x^2-3\gamma(\Lambda)\tilde{k}_y^2+\ldots
\end{equation}

\noindent where the subscript shows the cut-off dependence of $\Sigma(i\omega,\mathbf{k})$, while the coefficients $\alpha(\Lambda)$, $\beta(\Lambda)$, and $\gamma(\Lambda)$ are to be determined from \eqref{selfenergy}. Then, the Green's function reads

\begin{equation}\label{green}
G(\omega,\mathbf{k})=\left(G_0^{-1}(\omega,\mathbf{k})-\Sigma_\Lambda(\omega,\mathbf{k})\right)^{-1}=\dfrac{Z(\Lambda)}{\omega-Z_x^{-1}(\Lambda)k_x^2+3Z_y^{-1}(\Lambda)\tilde{k}_y^2+i\Gamma(\omega)}.
\end{equation}

\noindent Here we have absorbed all the divergencies into mass renormalization $Z_x(\Lambda)$ and $Z_y(\Lambda)$ as well as quasiparticle spectral weight $Z(\Lambda)$, whereas the imaginary part $\Gamma(\omega)$, that determines quasiparticle lifetime, will be unimportant for subsequent analysis. By requiring for $G(\omega,\mathbf{k})$ to be cut-off independent we derive a set of flow equations

\begin{equation}\label{cutoffindep}
\Lambda\dfrac{d}{d\Lambda}G^{-1}(\omega,\mathbf{k})=0
\end{equation}

\noindent Following the standard paradigm \cite{Salmhofer_sup,ZiJu_sup} and restricting to zero temperature regime we obtain

\begin{equation}
\Sigma_1(\omega,\mathbf{k})=\tilde{U}^2\left(A_1\omega+B_1k_x^2-3C_1\tilde{k}_y^2\right)\log^2\left(\dfrac{\Lambda}{2t}\right),
\end{equation}

\noindent and

\begin{equation}
\Sigma_{2,3}(\omega,\mathbf{k})=\tilde{U}^2\left(A_{2,3}\omega+B_{2,3}k_x^2-3C_{2,3}\tilde{k}_y^2\right)\log\left(\dfrac{\Lambda}{2t}\right),
\end{equation}

\noindent where $\tilde{U}=U/(2\pi^2t^2\sqrt{3})$. Finally, taking account of \eqref{cutoffindep} and factorizing $Z=Z_1Z_2Z_3$, as well as $Z_x$ and $Z_y$, we derive

\begin{align}
\label{rg}
\dfrac{d}{d\lambda}\log Z_1&=2g_4^2\lambda A_1, \\
\dfrac{d}{d\lambda}\log Z_2&=\left(g_1^2-g_1g_2+g_2^2\right)A_2, \\
\dfrac{d}{d\lambda}\log Z_3&=g_3^2 A_3.
\end{align}

\noindent Equations for $Z_x$ and $Z_y$ coincide with (\ref{rg}) provided $A_i$ is replaced by $A_i\rightarrow B_i-A_i$ and $A_i\rightarrow C_i-A_i$ respectively. The renormalized quasiparticle spectrum is determined by the poles of the Green's function \eqref{green} and is plotted in main text.

\section{Dual Fermion Spectral Functions}

In order to improve the stability of the Pad\'e analytical continuation within the dual fermion approach, we make use of improved estimators \cite{improvedest_sup} for the self-energy and vertex functions in the continuous-time hybridization expansion quantum Monte Carlo impurity solver \cite{cthyb_sup} (the calculations are free of the sign problem). To obtain reliable spectra we calculated them by analytical continuation of the self-energy
\begin{align}
\Sigma_{\omega}(\kv) = \Sigma^{\text{DMFT}}_{\omega} + \Sigma^{\text{NL}}_{\omega}(\kv)
\end{align}
and also of the Green's function
\begin{align}
\label{glat_sup}
G_{\omega}(\kv)=[(g_{\omega}+g_{\omega} \tilde{\Sigma}_{\omega}(\kv)  g_{\omega} )^{-1}+\Delta _{\omega } -\varepsilon_\mathbf{k} ]^{-1}.
\end{align}
The statistics of the simulations was sufficiently high that the results of both methods were consistent. We additionally verified the equality of the DOS on the real axis obtained from analytical continuation of the momentum resolved as well as the local Green's function, which poses a rather strict consistency check on the Pad\'e data.\newline
\begin{figure}[htbp!]
	\begin{minipage}[t]{0.45\textwidth}
	\includegraphics[width=\textwidth]{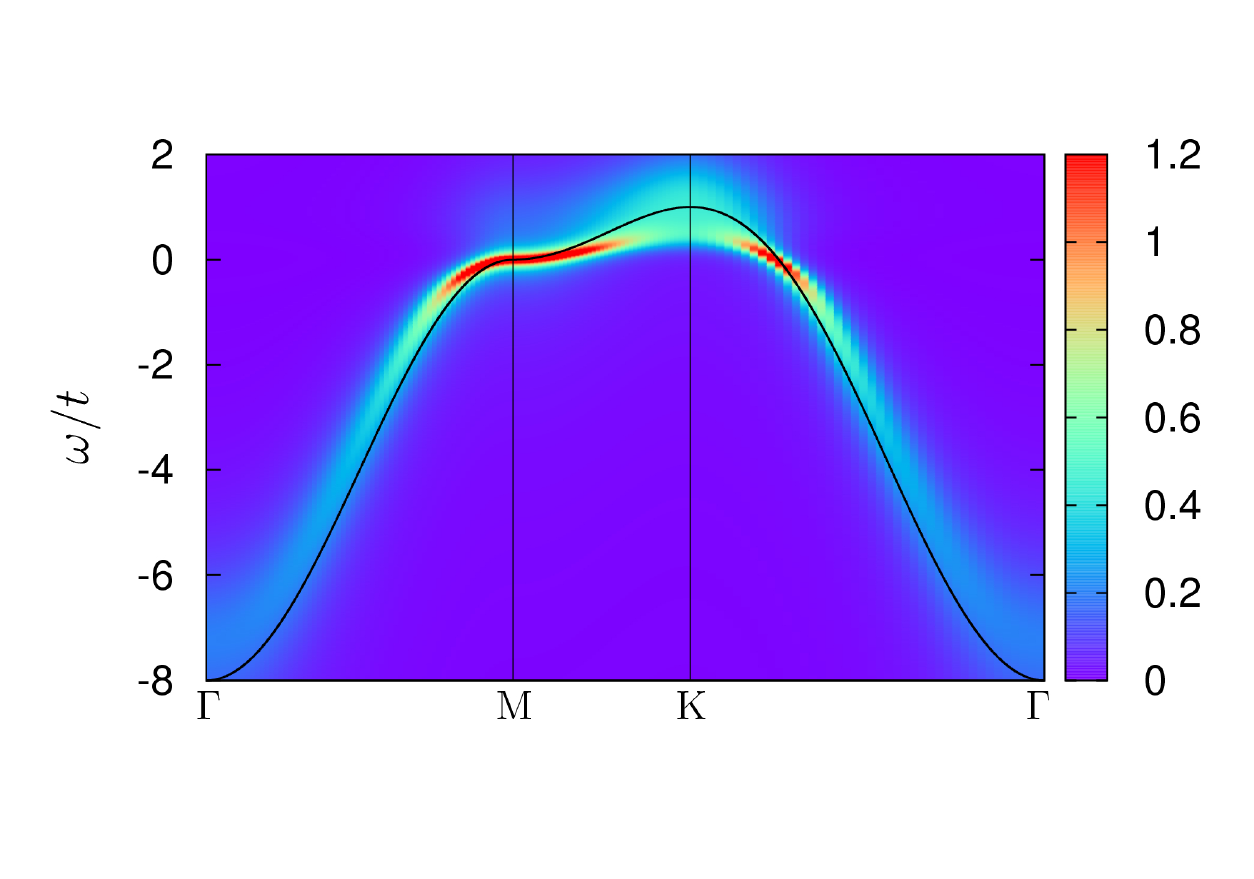}
	\end{minipage}
	\begin{minipage}[t]{0.45\textwidth}
	\includegraphics[width=\textwidth]{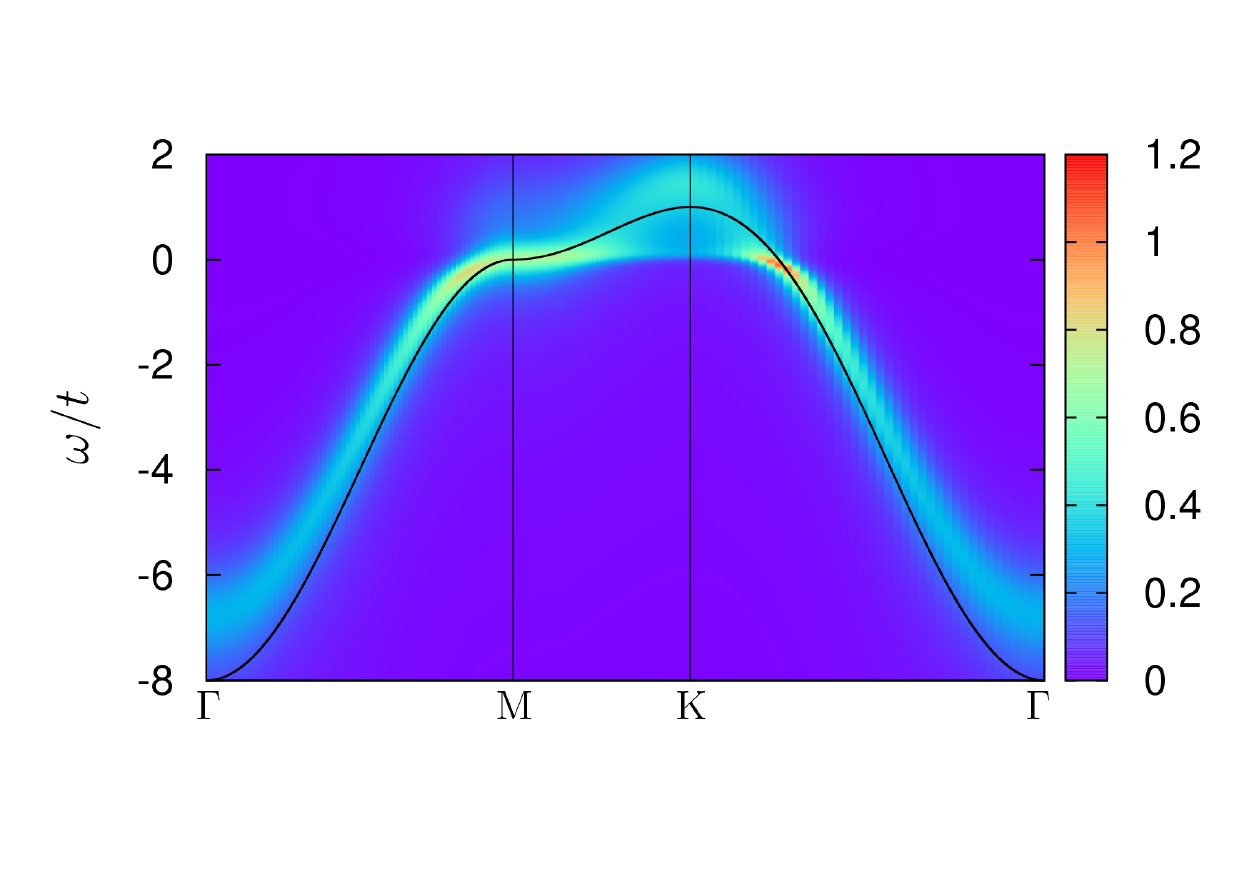}
	\end{minipage}
 \caption{Spectral functions for $U/t=8$ and $T/t=0.05$. \textit{Left panel:} Spectral function obtained within DMFT. \textit{Right panel:} Spectral function obtained from the full ladder sum of dual diagrams (LDFA), taking into account both charge and spin fluctuations. Flattening of the spectrum at the M-point can already be observed in DMFT to some degree, but the effect is significantly more pronounced in LDFA. One can see an extended VHS spanning a large region of the Brillouin zone. The effect is absent in second-order dual perturbation theory (not shown).}
 \label{fig:specfunc1}
\end{figure}

\begin{figure}[htbp!]
	\begin{minipage}[t]{0.45\textwidth}
	\includegraphics[width=\textwidth]{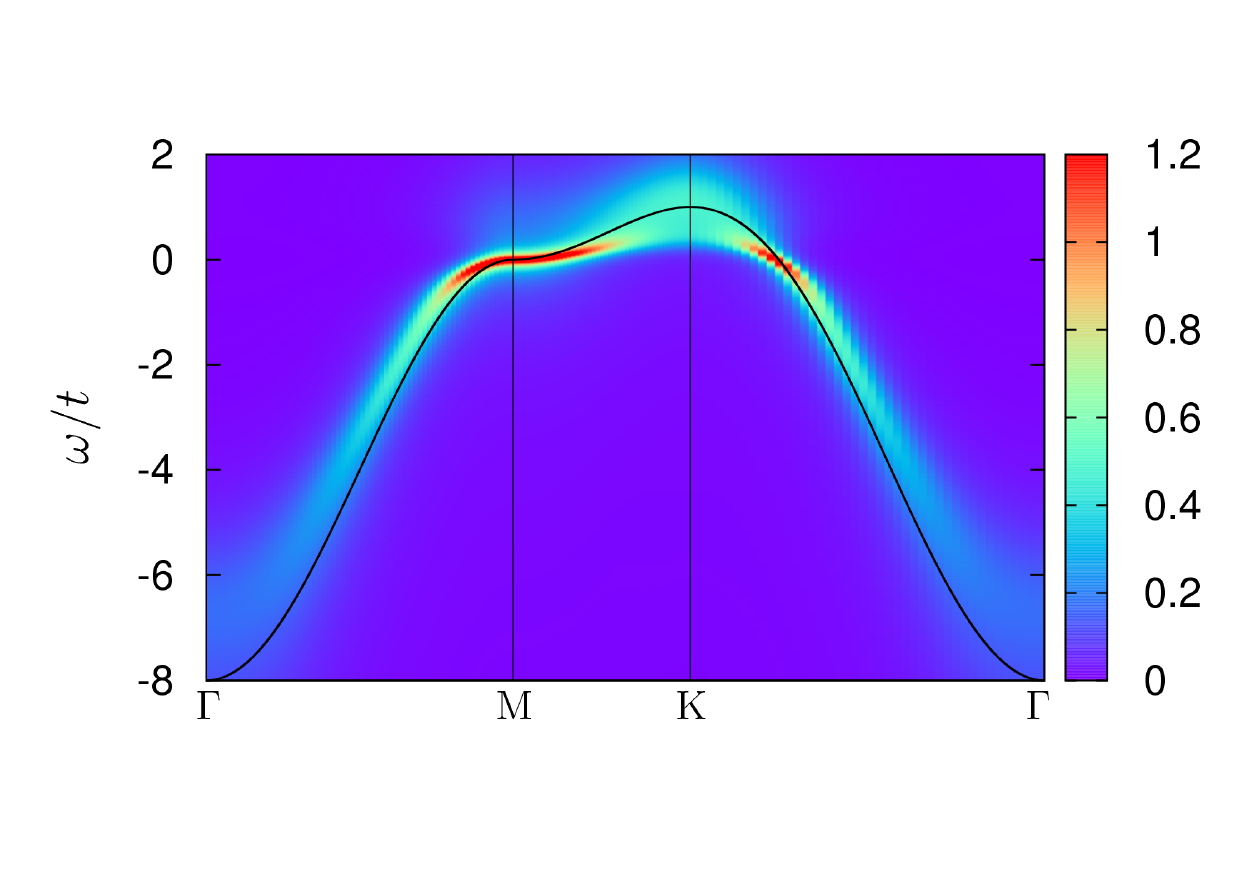}
	\end{minipage}
	\begin{minipage}[t]{0.45\textwidth}
	\includegraphics[width=\textwidth]{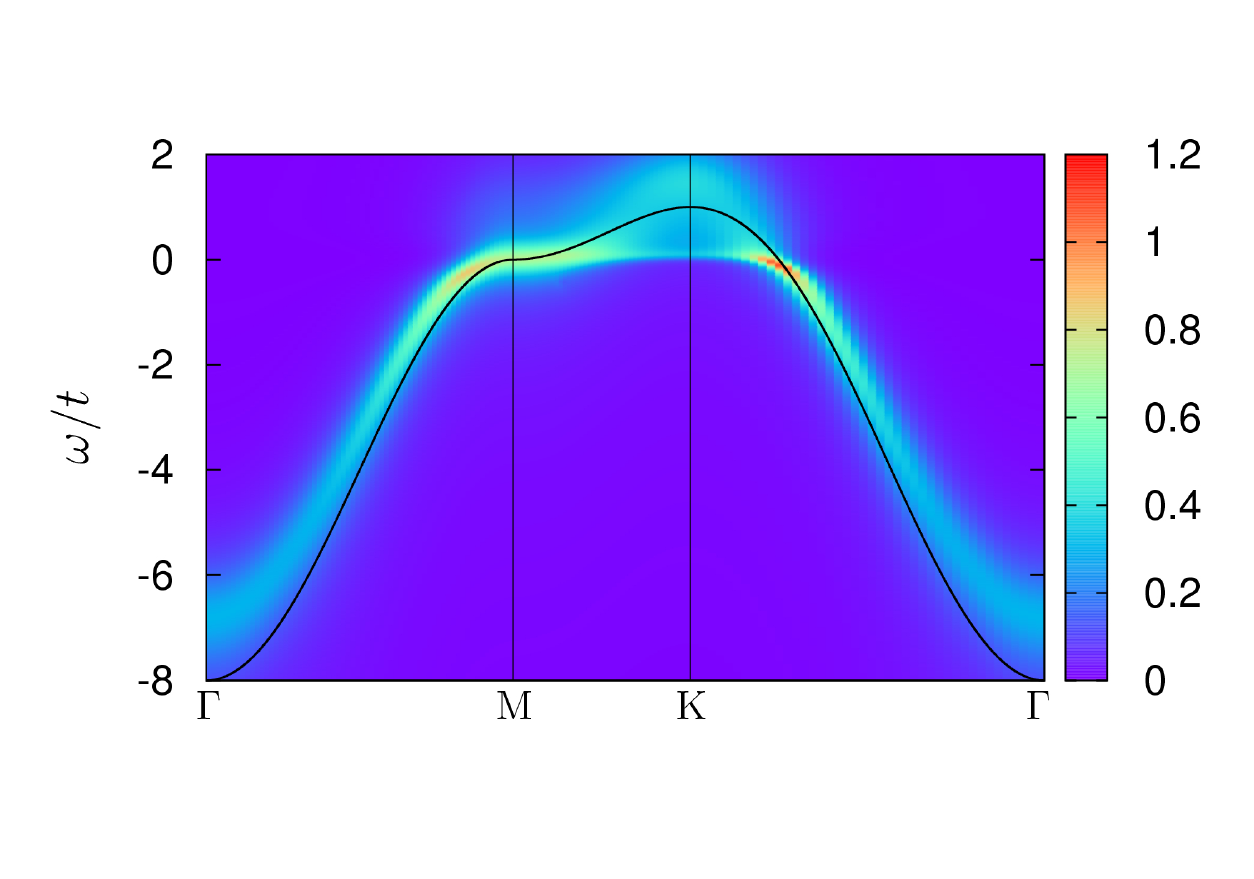}	
	\end{minipage}
 \caption{Spectral functions at $U/t=8$ and $T=0.05t$. \textit{Left panel:} Spectral function from charge LDFA. \textit{Right panel:} Spectral function from spin LDFA. The presence of the extended VHS in spin LDFA and its absence in charge LDFA shows that the effect is driven by collective spin excitations.
Note that the charge LDFA spectral function is very similar to the DMFT result, showing that corrections due to the charge channel are rather small.}
 \label{fig:specfunc2}
\end{figure}

\begin{figure}[htbp!]
	\begin{minipage}[t]{0.45\textwidth}
	\includegraphics[width=\textwidth]{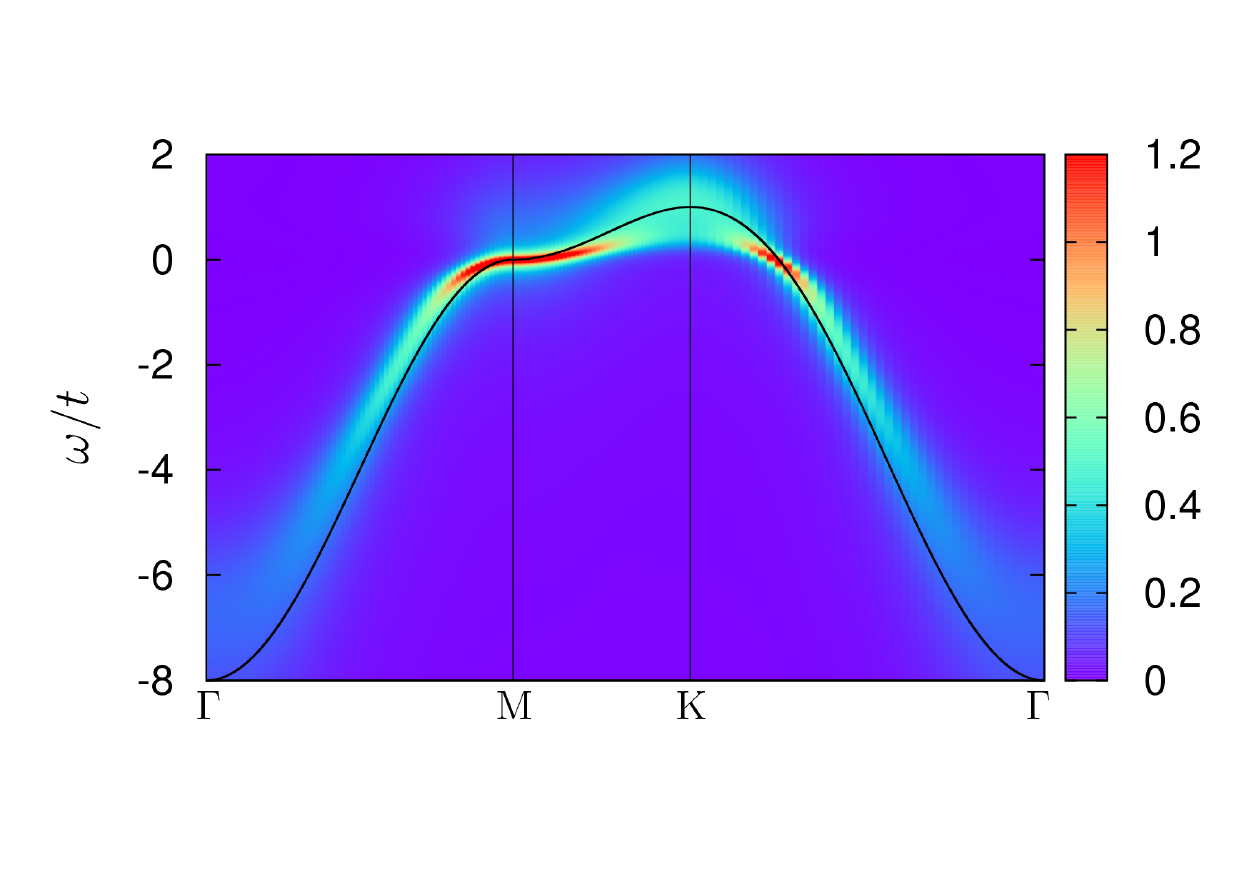}
	\end{minipage}
	\begin{minipage}[t]{0.45\textwidth}
	\includegraphics[width=\textwidth]{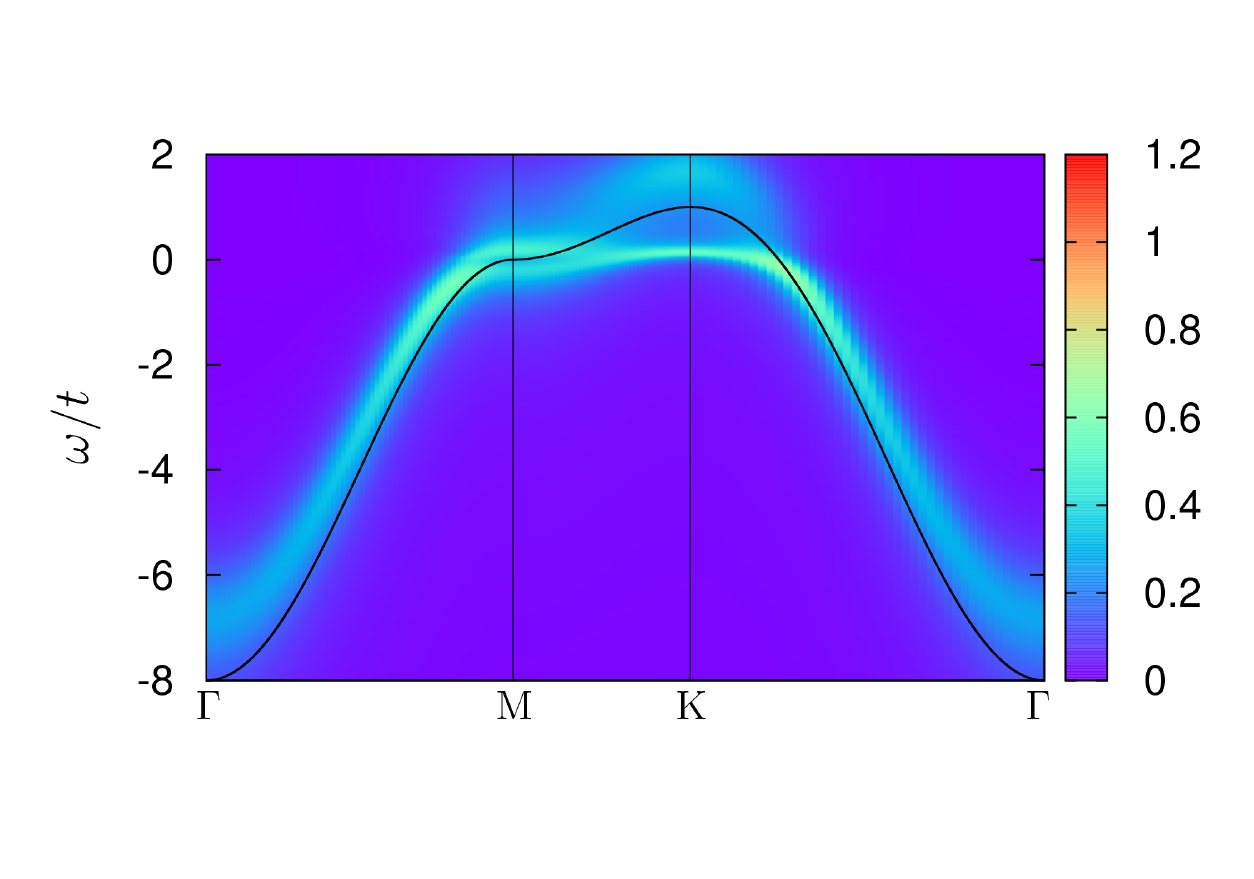}
	\end{minipage}
 \caption{Spectral functions for $U/t=8$ and $T/t=0.05$ without self-consistent renormalization of the Green's function. \textit{Left panel:} Spectral function obtained from non-self-consistent charge LDFA. \textit{Right panel:} Spectral function obtained from non-self-consistent spin LDFA. The band flattening in the non-self-consistent spin LDFA is significantly more pronounced (and overestimated) compared to the self-consistent calculations. At the M-point the dispersion is split.}
 \label{fig:specfunc3}
\end{figure}
The ladder dual fermion approach relies on the numerical solution of the Bethe-Salpeter equation, with a local approximation to the irreducible dual fermion vertex $\Gamma_\text{irr}\approx\gamma^{(4)}$. Here $\gamma^{(4)}$ plays the role of the bare interaction of the dual fermions, which is given by the fully antisymmetric, reducible two-particle vertex of the impurity problem. Solution of the BSE yields an approximation to the full vertex and subsequent application of the Schwinger-Dyson equation yields the dual self-energy in the ladder dual fermion approximation (LDFA),
\begin{align}
\tilde{\Sigma}_{\omega}(\kv) =\ \ T\frac{1}{2}&\sum_{\KV\omega'\nu} \gamma^{\nu\,\text{ch}}_{\omega\omega'} \tilde{G}_{\omega+\nu}(\kv+\KV)\tilde{\chi}^\nu_{\omega'}(\KV)[\Gamma^{\nu\,\text{ch}}_{\omega'\omega}(\KV)-\frac{1}{2}\gamma^{\nu\,\text{ch}}_{\omega'\omega}]\notag\\
+T \frac{3}{2}&\sum_{\KV\omega'\nu} \gamma^{\nu\,\text{sp}}_{\omega\omega'} \tilde{G}_{\omega+\nu}(\kv+\KV)\tilde{\chi}^\nu_{\omega'}(\KV)[\Gamma^{\nu\,\text{sp}}_{\omega'\omega}(\KV)-\frac{1}{2}\gamma^{\nu\,\text{sp}}_{\omega'\omega}],
\end{align}
where we have written spin and charge contributions explicitly. This allows us to investigate their respective influence on the quasiparticle spectrum. In the following, calculations considering either only the charge or only the spin channel of the BSE will be referred to as charge and spin LDFA, respectively. Figure \ref{fig:specfunc1} shows spectral functions obtained from DMFT and (full) LDFA. While band flattening is observed in DMFT to some extent, the formation of the extended VHS is due to the effect of spatial correlations.

In Fig. \ref{fig:specfunc2} we compare results from charge and spin LDFA. The formation of the extended VHS is evident in the spin LDFA, while it is absent in charge LDFA. One can further observe that the spectral weight at the M-point is significantly reduced as a result of interaction of quasiparticles with collective spin-excitations with corresponding momentum. From the similarity of the DMFT and charge LDFA result we conclude that the feedback from nonlocal charge excitations onto the self-energy (and thus the quasiparticle spectrum) is small.

The effect is even more pronounced where the ladder diagrams are evaluated with bare dual Green's functions instead of summing diagrams with renormalized propagators. This can be seen in Fig. \ref{fig:specfunc3} where the corresponding spectral functions for non-self-consistent charge and spin LDFA are shown. The spin LDFA spectral function even exhibits a splitting of the band at the Fermi level and in the vicinity of the VHS.
Note that while this calculation shows the formation of the extended VHS more clearly, it overestimates the effect: The strong attenuation of spectral weight at the Fermi level  diminishes the susceptibilities, which in turn leads to a smaller self-energy, etc. This effect is accounted for in the self-consistent calculation.

\begin{figure}[htbp!]
	\begin{minipage}[t]{0.45\textwidth}
	\includegraphics[width=\textwidth]{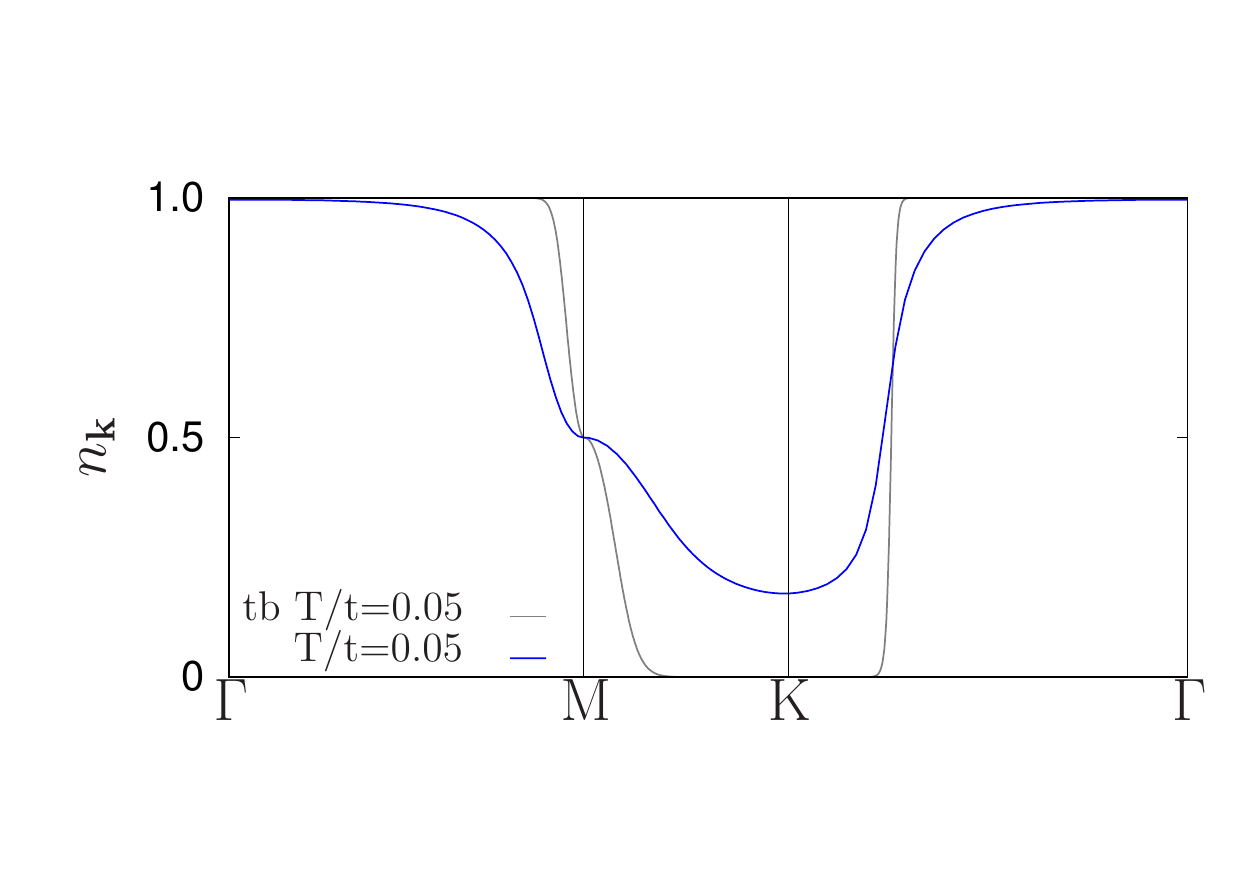}
	\caption{Occupation function plotted along the high symmetry path in momentum space as obtained from LDFA for U/t=8 and T/t=0.05 in comparison to the occupation function as obtained from tight-binding (tb) for the same temperature, showing the flattening at the M-point}
	\label{fig:nkU8vstb}
	\end{minipage}
	\hfill
	\begin{minipage}[t]{0.45\textwidth}
	\includegraphics[width=\textwidth]{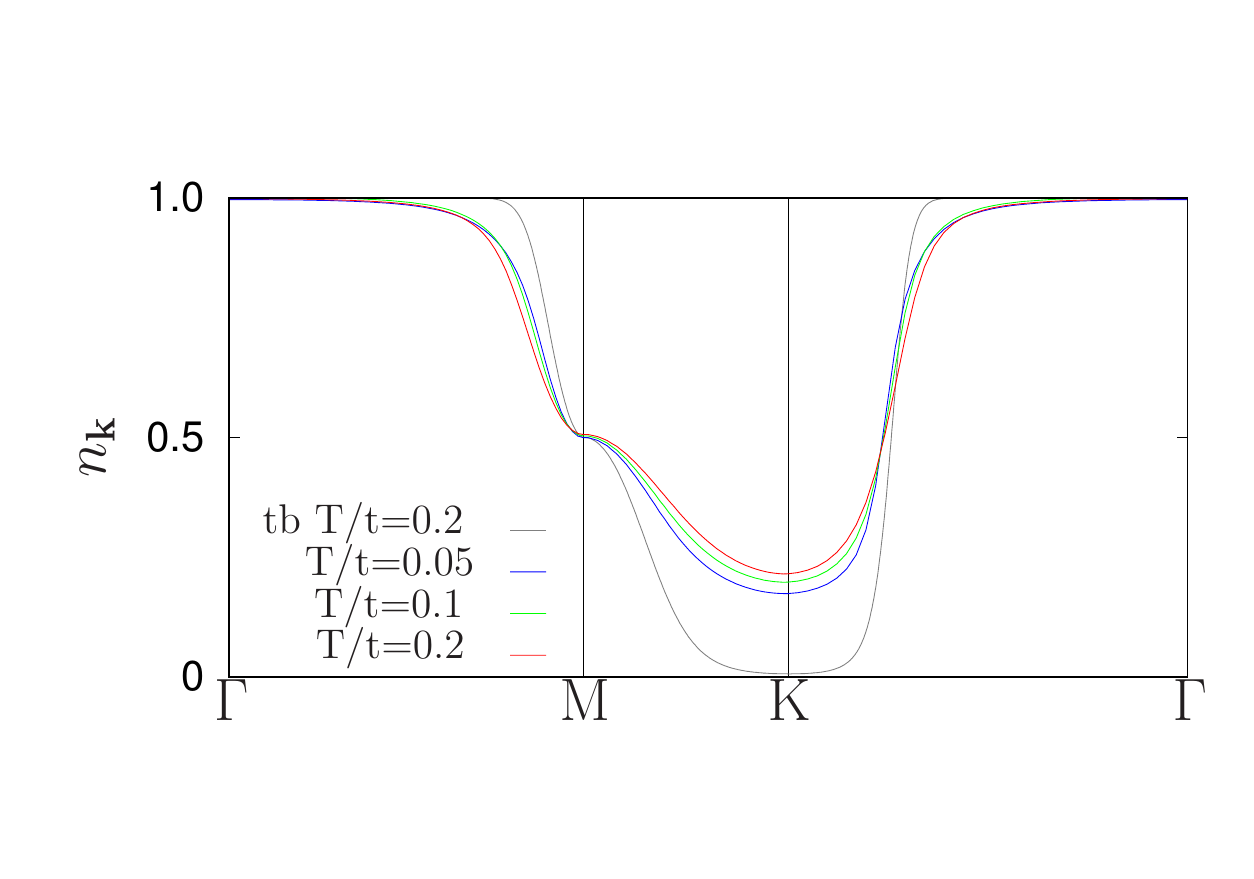}
	\caption{Occupation function plotted along the high symmetry path in momentum space as obtained from LDFA for U/t=8 for different temperatures, showing 					that the effect is robust under change in temperature. The occupation function obtained from tight-binding (tb) for T/t=0.2 is also shown as a 			reference. }
	\label{fig:nkU8vstb_T}
	\end{minipage}
\end{figure}
\begin{figure}[htbp!]
	\begin{minipage}[t]{0.33\textwidth}
	\includegraphics[width=\textwidth]{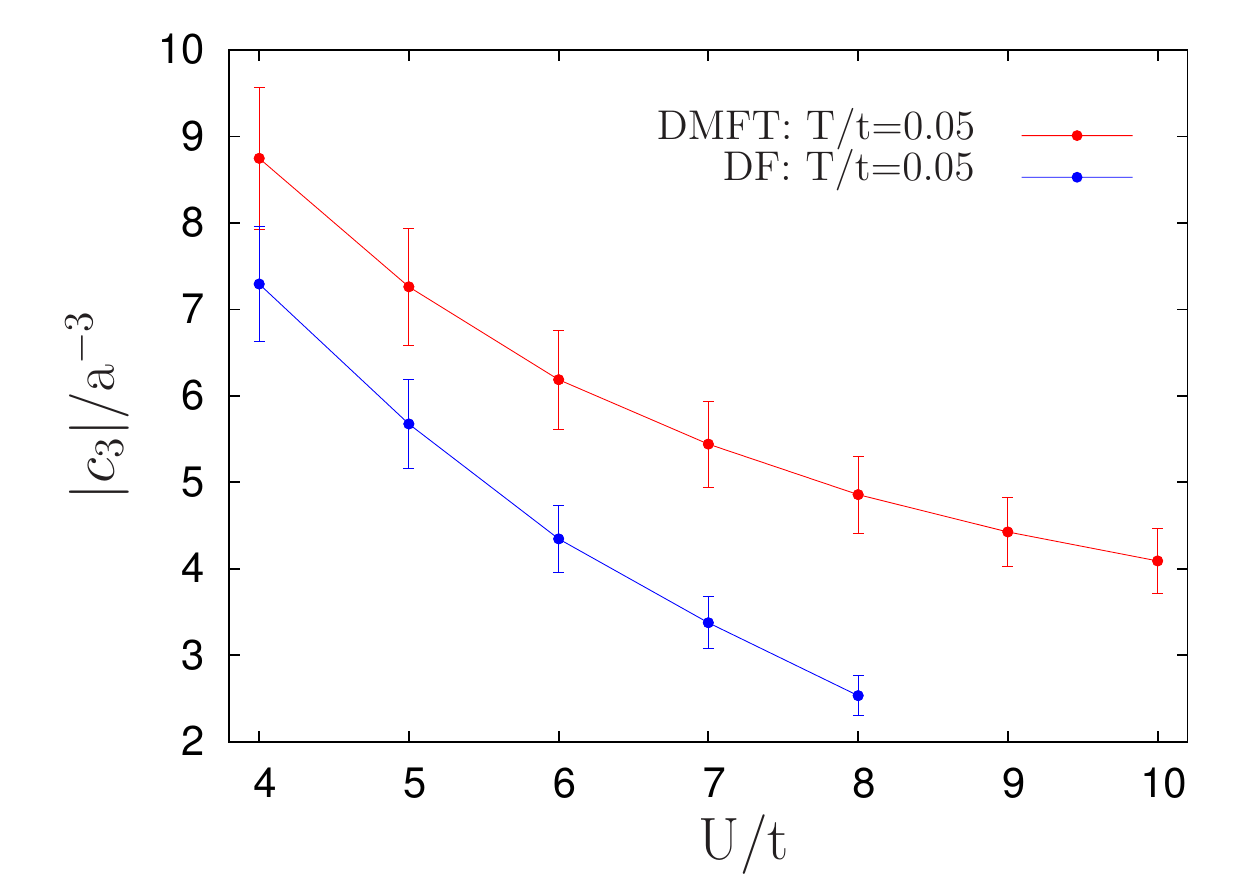}
	\end{minipage}
	\begin{minipage}[t]{0.33\textwidth}
	\includegraphics[width=\textwidth]{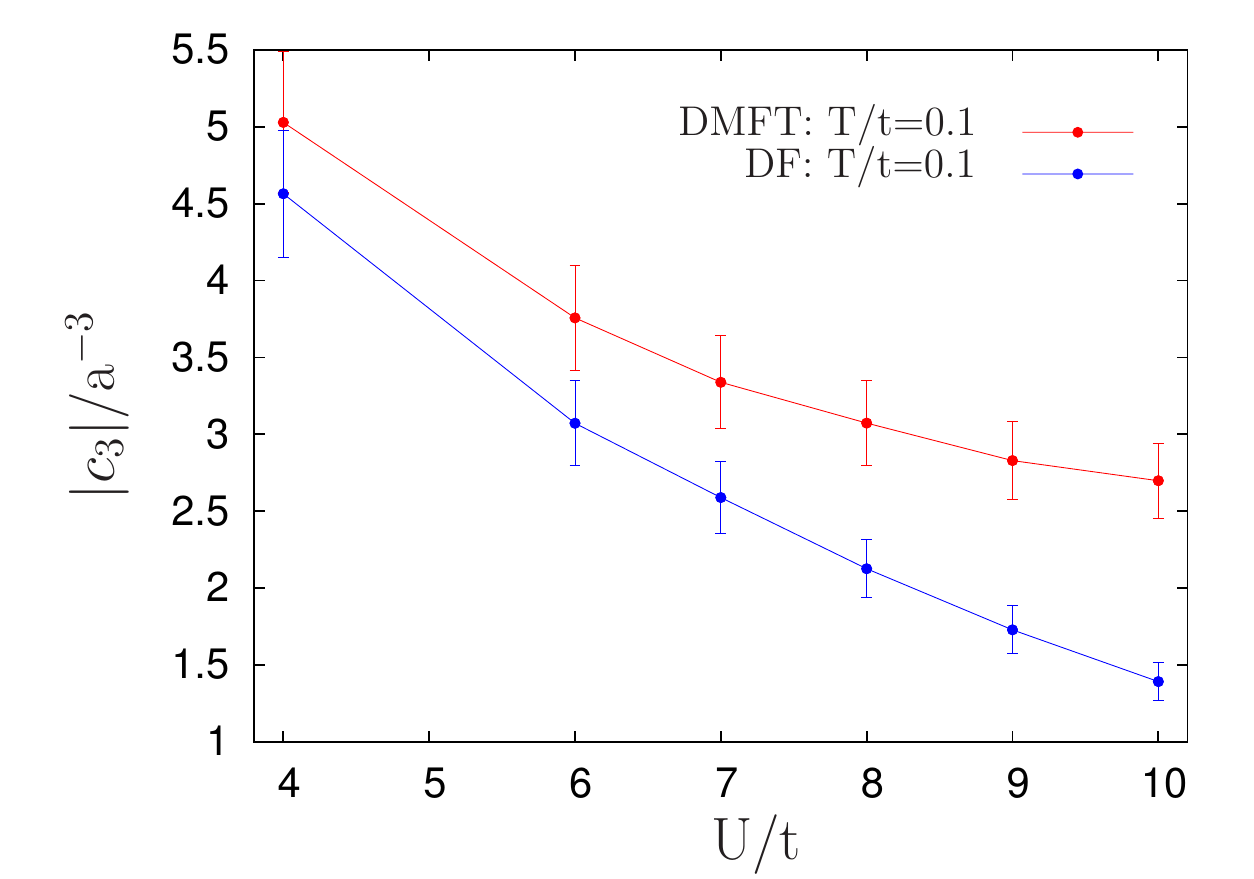}
	\end{minipage}
	\begin{minipage}[t]{0.33\textwidth}
	\includegraphics[width=\textwidth]{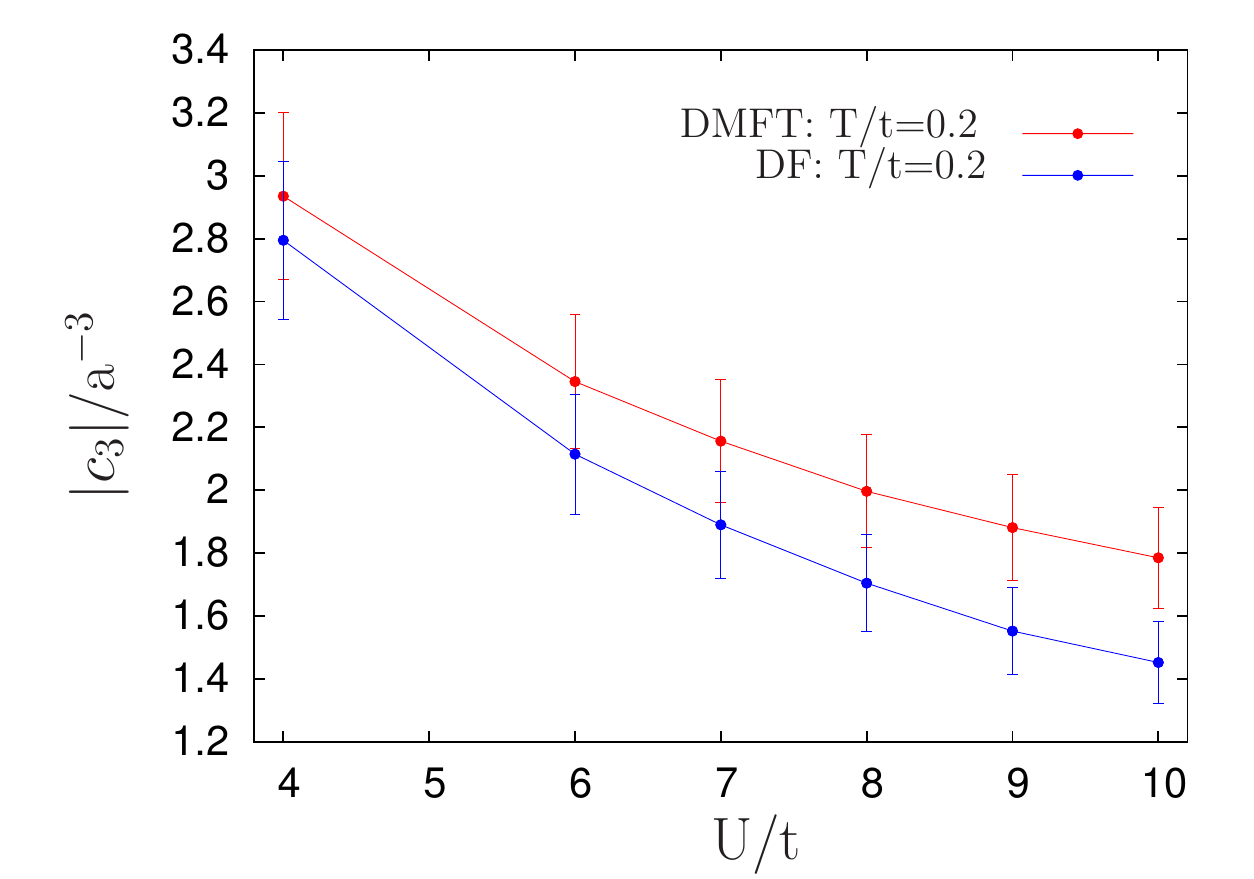}
	\end{minipage}
 \caption{Magnitude of the coefficient in front of the third order term of a third degree polynomial, obtained from a fit to the occupation function within a small interval around the M-point. Fits have been performed to occupation functions obtained from DMFT and LDFA. \emph{Smaller} magnitude implies a band which is flatter. The flattening of the occupation function increases with the interaction and is considerably stronger when nonlocal correlations are taken into account in the dual fermion calculation. Although the flattening enhances (the magnitude of $c_{3}$ decreases) with increasing temperature (from left to right), the correlation driven effect can clearly be separated from the purely thermal effect: In the plotted interaction range, the magnitude of $c_{3}$ changes by a factor of $2$ even at the highest temperature.}
 \label{fig:fitparam}
\end{figure}
The flattening of the spectrum at the M-point directly manifests itself in the occupation function which is plotted in Figs. \ref{fig:nkU8vstb} and \ref{fig:nkU8vstb_T}. The effect is robust under changes in temperature as can be seen from Fig. \ref{fig:nkU8vstb_T} and can still be observed at T/t=0.2, which is accessible in experiments with ultra cold gases in optical lattices. The flattening becomes more pronounced as the interaction increases. In order to quantify this behavior we have fitted a third degree polynomial $f(x)=c_0+c_1 x+c_2 x^{2}+c_3 x^{3}$ to the occupation function in a small interval around the M-point. Increased flattening is indicated by a \emph{decrease} in magnitude of the coefficient $c_3$ in front of the third order term. One can see in Fig. \ref{fig:fitparam}, that the coefficient therefore decreases with increasing interaction, as expected.
The dual fermion results are systematically below the DMFT result. Hence the band flattening predicted by LDFA is clearly more pronounced due to the effect of nonlocal correlations on the quasiparticle spectrum.
With increasing temperature (from left to right) the differences become smaller since non-local correlations become weaker due to thermal fluctuations. The magnitude of the coefficient also decreases with increasing temperature (note the different scales).  This purely thermal effect which does not depend on the interaction. The corresponding flattening can be explained by the broadening of the Fermi function and can be seen by comparing occupation functions in the tight-binding approximation in Figs. \ref{fig:nkU8vstb} and \ref{fig:nkU8vstb_T}. The correlation-driven effect can be observed by tuning the on-site interaction $U$. Despite being weakened at higher temperatures, the interaction dependence remains strong and clearly visible up to the highest temperature.

\end{document}